\documentclass[5p,sort&compress]{elsarticle} 

\usepackage{amsmath}
\usepackage{nccmath}
\usepackage{amsthm}
\usepackage{amsfonts}
\usepackage{bm}                 
\usepackage{mathrsfs}
\usepackage{dsfont}

\def \D {\mathrm{\,d}}

\def \P {\partial}
\def \I {\mathrm{i}}
\def \E {\mathrm{e}}

\begin{document}
\title{On the applicability of the two-band model to describe transport across
  n-p junctions in bilayer graphene}

\date{\today}

\author{C.~J.~Poole}
\ead{c.poole@lancaster.ac.uk}
\address{Department of Physics, Lancaster University, Lancaster, LA1 4YB, UK}

\begin{abstract}
  We extend the low-energy effective two-band Hamiltonian for electrons in
  bilayer graphene (Ref. \cite{McCann:2006p75}) to include a spatially dependent
  electrostatic potential. We find that this Hamiltonian contains additional
  terms, as compared to the one used earlier in the analysis of electronic
  transport in n-p junctions in bilayers
  (Ref. \cite{Katsnelson:2006p58}). However, for potential steps $|u|<\gamma_1$
  (where $\gamma_1$ is the interlayer coupling), corrections to the transmission
  probability due to such terms are small. For the angle-dependent transmission
  $T(\theta)$ we find $T(\theta)\cong\sin^2(2 \theta)-(2 u/3 \gamma_1) \sin(4
  \theta) \sin(\theta)$ which slightly increases the Fano factor: $F \cong
  0.241$ for $u=40\mathrm{meV}$.
\end{abstract}


\maketitle

\begin{figure}
  \includegraphics[width=254pt]{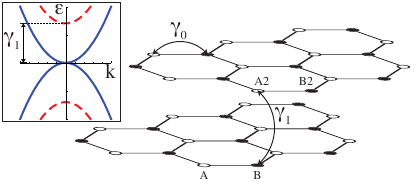}
  \caption{Schematic of AB (Bernal) stacked bilayer graphene showing intralayer and
    interlayer couplings, as well as a unit cell comprising of four carbon atoms:
    A,B,A2,B2. Inset: energy bands in bilayer graphene near a Kpoint. The energy
    of the quasiparticles is near $\varepsilon=0$, qualifying the assumption that
    $\gamma_1$ is large compared to other energies in the system. The
    transformation reduces the band structure to blue (solid) bands
    only.}\label{fig-bilayer-bond-schematic}
\end{figure}

Graphene, a crystal of carbon atoms in a two-dimensional (2D) honeycomb lattice, is a
gapless semiconductor \cite{Geim:2007p3540,McCann:2006p75}. Gating of graphene
enables one to vary the carrier density and therefore move the Fermi level from the
conductance band to the valence band. Gating graphene flakes with multiple gates
enables one to generate electrostatically defined n-p junctions
\cite{Cheianov:2006p82,Cheianov:2007p505,Katsnelson:2006p58, Snyman:2007p2060,
  Cayssol:2009p3671,Stander:2009p3675,Huard:2007p3680, Williams:2007p3790,
  Gorbachev:2008p4098, Cserti:2007p3978,Peterfalvi:2009p3977,
  Fogler:2008p4051,Zhang:2008p4052}. Bilayer graphene in particular is often
described by a four-band Hamiltonian from a tight-binding calculation (given
that there are four atoms in the unit cell; see
Fig. \ref{fig-bilayer-bond-schematic}). For low energies near the Fermi surface,
one can describe the transport of electrons with a two-band Hamiltonian
\cite{McCann:2006p75}. Transport across an n-p junction in bilayer graphene in
the low-energy, ballistic regime has been previously studied in
Ref. \cite{Katsnelson:2006p58}, but without considering the possibility of a
correction due to the spatial dependence of the electrostatic potential.

\begin{figure}
  \includegraphics[width=254pt]{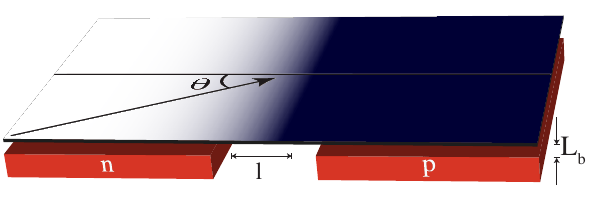}
  \caption{Angular dependence of quasiparticle transmission through an n-p junction.}\label{fig-np-angular-dependence}
\end{figure}

In this paper, we extend the derivation of an effective two-band Hamiltonian for
bilayer graphene (in the low-energy regime) to include the effects of a
spatially dependent electrostatic potential $u$, and a gap in the energy spectrum
$\Delta$. The re-derived two-band model Hamiltonian contains several additional
terms which originate from the spatial derivatives of $u(x)$. We use this in the
analysis of the problem of an n-p junction, where we find a change in
transmission probability, as compared to the analysis in
Ref. \cite{Katsnelson:2006p58}, which showed perfect transmission through the
n-p junction at an angle of $45^\circ$ (see
Fig. \ref{fig-np-angular-dependence}). This analysis shows that the additional terms
in the effective two-band Hamiltonian induced by the gradient expansion
involving the lateral potential are small, and thus the correctional term to
the angular transmission probability increases the angle at which perfect
transmission occurs by a few degrees. This also results in a small correction to
the Fano factor.

\begin{figure}
  \includegraphics[width=254pt]{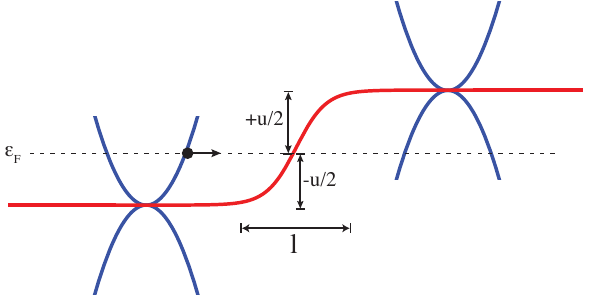}
  \caption{Low-energy band structure of a single valley on either side of the
    potential step. The Fermi energy is the same on both sides, causing an
    electron from the left side to tunnel through the barrier from the
    conductance band to the valence band on the right side.}\label{fig-np-junction-schematic}
\end{figure}

Using the nearest-neighbour tight-binding approximation in the
Slonczewski-Weiss-McClure parameterisation \cite{Slonczewski:1958p61}, one can
write the Hamiltonian at a K point (for basis
$(\phi_A,\phi_{B2},\phi_{A2},\phi_{B})$) as

\begin{equation}\label{eqn-bilayer-effective-Hamiltonian-initial-ham} {\cal
H}_\mathrm{4x4} = 
\begin{pmatrix}
- \xi\frac{\Delta}{2}\sigma_z+\hat u & \xi v\bm{\sigma}\cdot\bm{p}\\
\xi v\bm{\sigma}\cdot\bm{p} & \xi\frac{\Delta}{2}\sigma_z + \gamma_1
\sigma_x+\hat u
\end{pmatrix},
\end{equation}
where $\bm{\sigma}=(\sigma_x,\sigma_y)$, $\bm{p}=(p_x,p_y)$ and $\xi$ is the
Dirac point index ($\xi=+1$ for the valley around the $\mathrm{K}$ point, $-1$
for the valley around the $\mathrm{K'}$ point, and throughout this paper we set
$\hbar=1$). $v=\frac{\sqrt{3}}{2}a\gamma_0/\hbar$ and $\sigma_{i}$ are the Pauli
spin matrices. Furthermore, $\Delta=\varepsilon_2-\varepsilon_1$ is the
difference between the on-site energies in the two layers,
$\varepsilon_2=\frac{1}{2}\Delta$, $\varepsilon_1=-\frac{1}{2}\Delta$, which
produces a gap in the energy spectrum \cite{McCann:2006p68}. A potential term
$\hat{u}=\mathbb{I} u$ is added along the diagonal to represent the
electrostatic potential (we neglect inter-valley scattering between $\mathrm{K}$
and $\mathrm{K'}$; $\mathbb{I}$ is the unit matrix). We assume that the
interlayer coupling $\gamma_1$ is large compared to other energies in the system
(which is reasonable for the low-energy regime near the Dirac points). Given
that $\varepsilon\ll\gamma_1$ (where $\varepsilon$ is the energy of charge
carriers), and with $\varepsilon=p^2/2m$, where $m=\gamma_1/2v^2$, we see that
$(pv/\gamma_1)^2\ll1$. From this justification, we drop terms beyond quadratic
in momentum in the following calculations. We assume a non-adiabatic system,
with
\begin{equation}\label{eqn-length-scales}
a\ll l_\perp\ll (l, \lambda_F), 
\end{equation} 
where $a$ is the lattice constant, $l$ the width of the step (see
Fig. \ref{fig-np-junction-schematic}), $l_\perp=v/\gamma_1$, and $\lambda_F$ is
the Fermi wavelength.

We use a Schrieffer-Wolff transformation \cite{Schrieffer:1966p370} to map
Eq. (\ref{eqn-bilayer-effective-Hamiltonian-initial-ham}) in a 4D Hilbert space
into a 2D subspace, creating an effective Hamiltonian. If we let ${\cal
  H}_{\mathrm{4x4}}={\cal H}^0 + \delta{\cal H}$, with

\begin{equation} {\cal H}^0 =
\begin{pmatrix}
  H_{11} & 0\\
  0 & H_{22}
\end{pmatrix}, \qquad \delta{\cal H} = 
\begin{pmatrix}
  0 & H_{12}\\
  H_{21} & 0
\end{pmatrix},
\end{equation}
we can then write the associated Green's function as ${\cal
  G}_\mathrm{4x4}=(\varepsilon-{\cal H}_\mathrm{4x4})^{-1}=(\varepsilon-{\cal
  H}^0-\delta{\cal H})^{-1}$ and expand:

\begin{fleqn}
  \begin{eqnarray} {\cal G}_\mathrm{4x4}& = & (\varepsilon-{\cal H}^0)^{-1} +
    (\varepsilon-{\cal H}^0)^{-1}\delta{\cal
      H}(\varepsilon-{\cal H}^0)^{-1} +\nonumber\\
    && (\varepsilon-{\cal H}^0)^{-1}\delta{\cal H}(\varepsilon-{\cal
      H}^0)^{-1}\delta{\cal H}(\varepsilon-{\cal H}^0)^{-1} +
    \cdots.\label{eqn-bilayer-effective-Hamiltonian-g4x4}
\end{eqnarray}
\end{fleqn}

Given the basis that ${\cal H}_\mathrm{4x4}$ is constructed in, and that the
low-energy quasiparticle transport is directly from atom A to B2 in the bilayer
unit cell \cite{McCann:2006p75} (see Fig. \ref{fig-bilayer-bond-schematic}), we
wish to map ${\cal H}_\mathrm{4x4}$ onto the $H_{11}$ block matrix, using a
Schrieffer-Wolff transformation. This has the effect of only keeping terms with
an even number of $\delta{\cal H}$ components. The end result is that
$G_{11}^{-1} = \varepsilon-H_{11}-H_{12}(\varepsilon-H_{22})^{-1}H_{21}$.

During this projection, the orthonormality of the wavevectors has to be
preserved. To do this, we notice that $G_{11}^{-1}$ is an inverse Green's
function of the form 

\begin{eqnarray} 
G_{11}^{-1} &=& \varepsilon-H_{11}+\Omega+\varepsilon\beta,\nonumber\\
\beta &=& \frac{1}{\gamma_1^2}H_{12}H_{21},\nonumber\\
\Omega &=& \frac{\xi \frac{\Delta}{2}
}{\gamma_1^2}H_{12}\sigma_z H_{21} -
\frac{1}{\gamma_1^2}H_{12}\hat u H_{21}\nonumber\\
&&+\frac{1}{\gamma_1}H_{12}\sigma_x H_{21}.
\end{eqnarray}

We write an effective Schr\"odinger equation as
$\varepsilon(1+\beta)|\psi\rangle=(H_{11}-\Omega)|\psi\rangle$. We wish to
enforce $\langle\psi|1+\beta|\psi\rangle=|\psi|^2$ and
$\langle\varphi|\psi\rangle = 0$. Writing the wavefunction in terms of a new
wavefunction $|\tilde\psi\rangle$,
$|\psi\rangle=(1+\beta)^{-1/2}|\tilde\psi\rangle$. Inserting this result back
into the effective Schr\"odinger equation gives us
$H_{\mathrm{eff}}=(1+\beta)^{-1/2}(H_{11} - \Omega)(1+\beta)^{-1/2}$, which
after Taylor expanding around $\beta$ up to ${\cal O}(\beta^2)$ produces

\begin{equation} 
  H_{\mathrm{eff}} = H_{11} - \Omega
  - \left\{(H_{11}-\Omega),\frac{1}{2\gamma_1^2}H_{12}H_{21}\right\},
\end{equation}
where the curly braces denote the anticommutator. The effective Hamiltonian can
thus be calculated as

\begin{fleqn} 
\begin{eqnarray}\label{eqn-bilayer-effective-Hamiltonian-final-Hamiltonian}
  H_\mathrm{eff} = &-&\frac{1}{2m}
  \left[\sigma_x\left(-k_y^2-\P_x^2\right) + 2\I
    \sigma_y k_y\P_x\right]\nonumber\\
  &+&\xi\frac{\Delta
    v^2}{\gamma_1^2}|\bm{p}|^2\sigma_z - \xi\frac{\Delta}{2}\sigma_z + \hat{u} \nonumber\\
  &+&\frac{v^2}{2\gamma_1^2}\left[(\bm\nabla^2 u) + 2 \bm\sigma (\bm\nabla u
    \times \bm p)\right] \nonumber\\
&+& \frac{v^2\xi}{4\gamma_1^2}\big[2 \mathbb{I} (\bm \nabla\Delta) \times \bm p\nonumber\\
    &&- \sigma_z\left\{4\left((\bm\nabla\Delta) \cdot\bm \nabla + \Delta\bm
        \nabla^2\right) + (\bm \nabla^2\Delta)\right\}\big].
\end{eqnarray}
\end{fleqn}

The first two lines form the Hamiltonian found in Ref. \cite{McCann:2006p75}
(neglecting trigonal warping). The additional correctional terms arise from the
spatial dependence of $u$ and $\Delta$. Their derivation and the following
analysis represent the subject and result of this paper.


The effective Hamiltonian in
Eq. (\ref{eqn-bilayer-effective-Hamiltonian-final-Hamiltonian}) can be
simplified when $\lambda_F\gg l$. Terms with $|\bm{p}|^2\sim k_F^2$ can be
dropped, given the length scales in this regime and the de Broglie relation. Now
we wish to compare terms containing the potential $u$ and gap $\Delta$. To do
this, we follow a simplified scheme to that defined in
Ref. \cite{McCann:2006p68}, modelling the bilayer on a substrate as a parallel
plate capacitor.

Each layer of graphene has surface area $A$, and we take the dielectric
constants of the material between the back gate and layer 1, and the bilayer, to
be unity. Layer 1 has charge $Q=-n_1eA$, while layer 2 has charge $Q'=-n_2eA$,
where $n_1$ ($n_2$) is the density on layer 1 (2) (and $n=n_1+n_2$). The back
gate and layer 1 are separated by a distance $L_b$, while the two layers are
separated by a distance $c_0$. Applying a Gaussian surface around layer 1, the
magnitude of the electric field is $E=Q/\varepsilon_0 A$, where $\varepsilon_0$
is the permittivity of free space. The voltage due to this electric field is
thus $QL_b/\varepsilon_0A$. The electric potential energy due to the back gate
(thus, the potential $u$) is
$u=eQL_b/\varepsilon_0A=n_1e^2L_b/\varepsilon_0$. We assume that the electric
field from the back gate is screened poorly by layer 1, so applying the same
analysis to layer 2, we find that the magnitude of the electric field is
$E'=Q'/\varepsilon_0A$. The voltage produced by that electric field is
$V'=E'c_0$, so the electric potential energy between the graphene layers (i.e.,
the gap) is $\Delta=n_2e^2c_0/\varepsilon_0$. If we assume that the charge
density is evenly distributed between the layers, $n_1=n_2=n/2$, then
$u/\Delta=L_b/c_0$. With $c_0\sim0.3\mathrm{nm}$ and $L_b\sim 300\mathrm{nm}$,
we find that $u\gg\Delta$. By writing $H_\mathrm{eff}$ in the form
$H_\mathrm{eff}=\mathbb{I} A + \sigma_x B + \sigma_y C + \sigma_z D$, we compare
each term and keep only the largest one in each group $A,B,C,D$.  This produces
an approximate Hamiltonian,

\begin{eqnarray}\label{eqn-bilayer-happrox}
H_\mathrm{app} = &-&\frac{1}{2m}\left[\sigma_x\left(-k_y^2-\P_x^2\right) + 2\I
  \sigma_y k_y\P_x\right]\nonumber\\
&+&\mathbb{I}\frac{k_F^2}{2 m}\left[ u + \frac{v^2}{2
    \gamma_1^2}\eta\left(\P_x^2 u\right)\right] \nonumber\\
&+& \sigma_z\left[-\xi\frac{\Delta}{2}+ \eta\frac{v^2k_F^2}{2m\gamma_1^2}
  \left(\P_x u\right) k_y\right],
\end{eqnarray}
where $\eta\in \{0,1\}$ and highlights the correctional terms.


An n-p junction can be formed with two back gates, schematically shown in
Fig. \ref{fig-np-angular-dependence}. Each gate can independently create an
electrostatic potential over that region of bilayer graphene. Given our chosen
length scales in Eq. (\ref{eqn-length-scales}), we model the n-p junction as a
Heaviside step function $\Theta(x)-(1/2)$, with its derivative the Dirac delta
function. Thus, $u \approx (k_F^2/2m) [\Theta(x)-(1/2)]$, which also determines
all additional terms in
Eq. (\ref{eqn-bilayer-effective-Hamiltonian-final-Hamiltonian}).

We define the problem in terms of plane waves on the left-hand and right-hand sides of
the junction, $\psi_1$ and $\psi_2$ respectively:

\begin{eqnarray} 
\psi_1 &=& 
\begin{pmatrix} 
1\\a_2
\end{pmatrix} 
\E^{\I k_x x}+ b
\begin{pmatrix}
1\\b_2
\end{pmatrix}
\E^{-\I k_x x} + c
\begin{pmatrix}
1\\c_2
\end{pmatrix}
\E^{-\kappa x},\nonumber\\
\psi_2 &=& d
\begin{pmatrix}
1\\d_2
\end{pmatrix}
\E^{-\I k_x' x} + f
\begin{pmatrix}
1\\f_2
\end{pmatrix}
\E^{\kappa' x}.
\end{eqnarray}

The Hamiltonian in
Eq. (\ref{eqn-bilayer-effective-Hamiltonian-final-Hamiltonian}) has plane and
evanescent wave solutions, and the quasiparticles are chiral, such that when they
pass from the conductance band at the left of the interface to the valence band
at the right, $k_x$ changes sign \cite{Katsnelson:2006p58} (see
Fig. \ref{fig-np-junction-schematic}).

With the step defined to be at $x=0$, we integrate
Eq. (\ref{eqn-bilayer-happrox}) across it,
$\int^{0+\delta}_{0-\delta}(\varepsilon-H_\mathrm{app})\D x$ and take the limit
$\delta\rightarrow 0$. Matching the wavefunctions at either side of the junction
($\psi_1(0)=\psi_2(0)$), we obtain the boundary condition

\begin{eqnarray}\label{eqn-np-boundary-condition}
  0&=&-\left.\frac{\sigma_x}{2m}\left(\P_x\psi\right)\right|^{\psi_2(0)}_{\psi_1(0)}+
  \mathbb{I}\,\eta \frac{k_F^2 v^2}{4m\gamma_1^2}\left(\frac{\psi_1'(0)+\psi_2'(0)}{2}\right)\nonumber\\
  &&- \sigma_z \eta \frac{v^2k_F^2k_y}{2m\gamma_1^2}\psi(0),
\end{eqnarray}
where the Fermi momentum $k_F=\sqrt{k_x^2+k_y^2}$ and

\begin{eqnarray}
a_2 &=& \frac{1}{\varepsilon-(u/2)}\left(\frac{k_y^2}{2m}-\frac{k_x^2}{2m}+\frac{\I k_x
  k_y}{m}\right),\nonumber\\
b_2 &=& a_2^*,\nonumber\\
c_2 &=& \frac{1}{\varepsilon-(u/2)}\left(\frac{k_y^2}{2m}+\frac{\kappa^2}{2m}-\frac{\kappa
  k_y}{m}\right),\nonumber\\
d_2 &=& \frac{1}{\varepsilon+(u/2)}\left(\frac{k_y^2}{2m}-\frac{k_x'^2}{2m}-\frac{\I k_x'
  k_y}{m}\right),\nonumber\\
f_2 &=& \frac{1}{\varepsilon+(u/2)}\left(\frac{k_y^2}{2m}+\frac{\kappa'^2}{2m}+\frac{\kappa'
  k_y}{m}\right).
\end{eqnarray}

\begin{figure}
  \includegraphics[width=254pt]{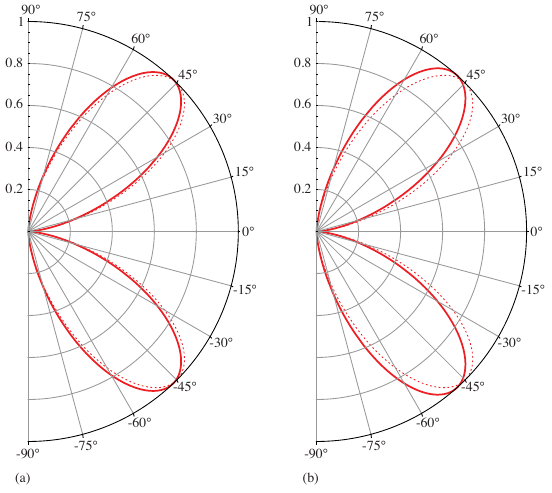}
  \caption{The dashed line shows the transmission probability without correctional terms
    applied. The solid line includes correctional terms. The dashed line shows perfect
    transmission at an angle of $45^\circ$ to the interface. (a) Transmission at
    $u=40\mathrm{meV}$. (b) Transmission at $u=80\mathrm{meV}$. Plotted with
    $\varepsilon=0$, $v\approx c/300$, $\gamma_1=0.4\mathrm{eV}$,
    $m=0.035\mathrm{m_e}$, $\Delta=0$.} \label{fig-transmission-2-plots}
\end{figure}

Using these equations, where $k_x=\sqrt{-k_y^2+2m[(u/2)-\varepsilon]}$,
$k_x'=\sqrt{-k_y^2+2m[(u/2)+\varepsilon]}$,
$\kappa=\sqrt{k_y^2+2m[(u/2)-\varepsilon]}$, and
$\kappa'=-\sqrt{k_y^2+2m[(u/2)+\varepsilon]}$, we calculate the transmission
probability for a symmetric junction $T(k_y)=|d|^2$. We assume a wide strip,
such that $k_y$ is invariant. We also set $\varepsilon=0$ in the middle of the
barrier for simplicity. Using $k_y = k_F \sin(\theta)$ (see
Fig. \ref{fig-np-angular-dependence}) we first calculate the transmission with only
the leading-order terms in Eq. (\ref{eqn-np-boundary-condition}) (by setting
$\eta=0$), finding agreement with Ref. \cite{Katsnelson:2006p58} in that
$T(\theta)=\sin^2(2 \theta)$. Including the correctional terms from
Eq. (\ref{eqn-np-boundary-condition}) by setting $\eta=1$, we obtain a
correction to the incident angle at which perfect transmission is seen (see
Fig. \ref{fig-transmission-2-plots}).

Taylor expanding the full analytical result for $T(\theta)$ around $\eta$, we
find that only the first-order term is important and obtain a
potential-dependent result (providing a good fit up to
$u\approx50\mathrm{meV}$):

\begin{equation}\label{eqn-junction-tfinal}
T(\theta) \cong \sin^2(2 \theta) - \frac{2 u}{3 \gamma_1}\sin(4 \theta)\sin(\theta).
\end{equation}


Assuming a wide graphene sheet (that is, a width $w$ much greater than the
length) and coherent quasiparticles, one can calculate the conductance from the
transmission probability using the Landauer-B\"uttiker approach
\cite{Buttiker:1986p3480} (taking into account two valleys and two spins),

\begin{equation}
G = \frac{4 e^2}{h}\sum_n|t_n|^2.
\end{equation}

\begin{figure}
  \includegraphics[width=254pt]{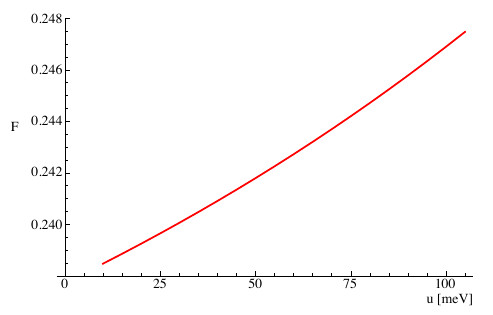}
  \caption{The Fano factor as a function of $u$, from a numerical calculation of the
    transmission probability for the same parameters as given in
    Fig. \ref{fig-transmission-2-plots}.} \label{fig-conductance-fano}
\end{figure}

With $k_y=2\pi n/w$ (where $n$ is an integer), we can write this as an integral
and calculate it using the full numerical transmission probability,

\begin{equation}
G = \frac{4 e^2 w k_F}{2 \pi h}\int^{\pi/2}_{-\pi/2}\D \theta \cos(\theta)
T(\theta) \cong2.1\frac{e^2 w
  k_F}{\pi h}
\end{equation}
for $u=40\mathrm{meV}$. This is a slight reduction from $2.12e^2 w k_F/\pi h$
for the case $\eta=0$. One can also calculate the Fano factor
\cite{FANO:1947p3476,Tworzydlo:2006p3485} (the ratio of shot noise to Poisson
noise; for a review see Ref. \cite{Blanter:2000p3322}) numerically:

\begin{equation} 
  F = \frac{\int_{-\pi/2}^{\pi/2} \D \theta\cos(\theta)
    T(\theta)(1-T(\theta))}{\int_{-\pi/2}^{\pi/2} \D \theta\cos(\theta)
    T(\theta)} \cong 0.241
\end{equation} 
for $u=40\mathrm{meV}$, showing a small increase compared to $F\cong0.238$ when
$\eta=0$ (see Fig. \ref{fig-conductance-fano}).

In conclusion, we have extended the earlier derived low-energy effective
Hamiltonian for bilayer graphene to incorporate a spatially dependent
electrostatic potential consistently. We calculate the angle-dependent
transmission through an n-p junction and find $T(\theta)\cong\sin^2(2 \theta)-(2
u/3 \gamma_1) \sin(4 \theta) \sin(\theta)$. Perfect transmission is still seen,
but at a slightly increased angle. The conductance is slightly reduced to
$G\cong2.1 e^2 w k_F/\pi h$, whereas the Fano factor is slightly increased to
$F \cong 0.241$ (both for $u=40\mathrm{meV}$).

The author thanks V.~I.~Fal'ko and V.~Cheianov for supervision during this
project and H.~Schomerus, E.~McCann and J.~Cserti for useful discussions. The
author thanks Lancaster University for financial support.



\end{document}